\documentclass{b8xegecls}
\twocolumn

\lat
\nonstopmode

\title{Solar neutrino background in  neutrinoless
       double beta-decay searching for  experiments}

\rtitle{Solar neutrino background in\ldots}

\sodtitle{Solar neutrino background in neutrinoless
          double beta-decay searching for experiments}

\author{Alexander \,A.\,Klimenko$^{+*}$
\/\thanks{e-mail: klimenko@nusun.jinr.ru}}

\rauthor{A.\,A.\,Klimenko }

\sodauthor{Klimenko~A.\,A.}

\address{$^+$Joint Institute for Nuclear Research, 141980 Moscow region,
 Russia\\~\\
$^*$Institute for Nuclear Research of RAS, 117312 Moscow,  Russia}

\abstract{Background of germanium and xenon detectors due to the
elastic scattering of  $^{8}B$ solar neutrinos on electrons have
been calculated. The background is 4.6 $\times 10^{-4}$ counts/(keV$\cdot t\cdot$y )
that defines the sensitivity of double beta-decay
experiments at a level $T^{0\nu\beta\beta}_{1/2} \approx
4\times10^{27}$~{y} and it corresponds to effective majorana
mass $ \mid < m_{\nu_e}>\mid \approx 23.5$ \ meV. This limit after
a half of one year of measurements for 10 tons natural Xe
detector will be obtained. }


\begin{document}

\maketitle

Nowadays neutrino is known to have  non zero rest mass.

Super-Kamiokande~\cite{superk}, SNO~\cite{sno},
KamLAND~\cite{kamland}  and radiochemical solar neutrino
experiments~\cite{sage},~\cite{gallexgno},~\cite{homestake} give
direct confirmation of neutrino oscillation and as a consequence
the existence of  non zero neutrino rest mass. Such type of
experiments yields information on the differences for squared
neutrino mass eigenvalues and mixing angles.

Neutrinoless double beta decay ( $0\nu\beta\beta$ )

\begin{equation}
  ^{A}_{Z}X \to ^{A}_{Z+2}X + 2e^{-}
\end{equation}

is characterized as a sensitive tool used both in physics beyond
the standard model and in reconstruction of the  neutrino mass
spectrum.

To obtain the absolute  neutrino mass estimations next-generation
of $0\nu\beta\beta$ experiments at large scale detectors for rare
events (neutrinoless double beta decay, dark matter and low energy
solar neutrino real time measurements), have been
proposed~\cite{genius,majorana,exo,xmass}, such as GENIUS, MAJORANA, EXO and XMASS.

These experiments are based on the search for the processes

\begin{equation}
\label{ge76}
  ^{76}Ge \to ^{76}Se + 2e^{-} , ^{136}Xe \to ^{136}Ba + 2e^{-}
\end{equation}

It is proposed to obtain a limit on the effective  Majorana  mass
$m_{\nu} \approx $15 \  meV, which corresponds to
$T_{1/2}^{0\nu}(0^+_i \to 0^+_f) \approx 10^{28}$y, while the
lowest current one is $ \mid < m_{\nu_e}>\mid $  < 350 meV ~\cite{klap}.

The effective Majorana mass of neutrino defines half-life of the
process~(\ref{ge76}) without contributions from right-handed
currents~\cite{doikot}

\begin{equation}
\label{halflife} T_{1/2}^{0\nu}(0^+_i \to 0^+_f)^{-1} =
\frac{{{\mid <m_{\nu}> \mid}^2}}{m_e^2} \times G^{0\nu}\times
M_{0\nu}^2
\end{equation}

where

 $$
\mid  < m_{\nu}> \mid  = {\mid U_{e1}\mid}^2 \ m_{1} +
 {\mid U_{e2} \mid}^2 \sqrt{\Delta m^2_{21}+ m_{1}^{2}} \  e^{i\phi_{2}} $$
$$
+ {\mid U_{e3}\mid}^2 \sqrt{ \Delta m^2_{32}+ \Delta m^2_{21}+m_{1}^{2}} \ e^{i\phi_{3}}
$$

is the effective majorana mass of neutrino, closely related with
parameters of neutrino oscillation experiments. Here
 $M_{0\nu}$ is $(0\nu\beta\beta)$ nuclear matrix element,
which can be calculated, while $U_{ei}$ are elements of the
neutrino mixing matrix, and
   $\Delta m^2_{ij} = m^2_i-m^2_j$,
where $m_i$ are neutrino mass eigenstates,
$\phi_i$ denote relative Majorana phases connected with CP
violation, and $G^{0\nu}$ is the phase space integral. The
experimental signature of the neutrinoless decay of $^{76}$Ge is a
peak at $Q_{\beta\beta}(^{76}Ge)=2039.006(50)$\
keV~\cite{douysset} and this value together with full width at
half maximum ( FWHM ) defines region of interest ( ROI~ ) of these
experiments. The background for these experiments should be at a
level of 0.1 counts /(keV$\cdot t\cdot$y) (the best present one is 60
counts/(keV$\cdot t\cdot$y)~\cite{klapd47kg}). For $^{136}Xe$ we have
$Q_{\beta\beta}(^{136}Xe)$= 2480  keV and FWHM $\approx$ 60 keV in ROI as
was declared in EXO and XMASS proposals.

 Solar neutrino flux and two
neutrino double decay mode represent unremovable background
components in such type of experiments making the recoil electron
signal undistinguishable from the two electron one in a germanium or xenon
detector.

The recoil electron energy of the elastic scattering ( ES ) process

\begin{equation}
                       \nu + e^- \rightarrow \nu + e^-
\end{equation}

is detected through ionization produced in the detector for searching for neutrinoless double beta decay.
 $^{8}B$ and hep neutrinos with energy up to 15 MeV and 18.8 MeV respectively can give background events into ROI.
Flux value of hep neutrinos in comparison with $^{8}B$ is
negligible and it will not be used in our analysis. Background
from $^{76}Ge(\nu,e^{-})^{76}As$ reaction, which is inverse beta
decay, is negligible due to nuclear level characteristics of
$^{76}As$ nucleus. For calculations of count rate in the region of
interest standard $^{8}$B solar neutrino spectrum was taken
from~\cite{bahcall}. Differential neutrino electron elastic
scattering cross section  was taken from~\cite{okun}.

\begin{equation}
\frac{d\sigma}{dT} = \frac{2G^{2} m_{e} }{ \pi }
\left[g_l^2+g_R^2(1-\frac{T}{E_{\nu}})^2-g_l g_R\frac{m_e
T}{E_{\nu}} \right]
\end{equation}

where $g_l =  \pm \frac{1}{2}+sin^2\theta_{W}$;
$g_R= sin^{2}\theta_{W}, sin^{2}\theta_{W}=0.23117$; with sign plus
corresponding to $\nu_e$, and sign minus to $\nu_{\mu,\tau}$;  G is the
Fermi constant. Cross section was calculated as

\begin{equation}
\sigma_{\nu{}_i}(^8B) = \int_0^{E_{\nu}^{MAX}} \Biggl(
\int_0^{T_{max}}  \frac{d\sigma_{\nu_i}(T,E_{\nu})}{dT} dT \Biggr)
\phi^{B-8}(E_{\nu}) dE_{\nu}
\end{equation}

where$ T_{max} = \frac{E_{\nu}}{1+m_e/2\times E_{\nu}}$ is maximal
kinetic energy of a recoil electron for neutrino with energy
$E_{\nu}$, and $i=e,\mu,\tau$.

Using $^8B$ neutrino fluxes measured in SNO experiment~\cite{sno}

$\phi^{CC}(\nu_e)=( 1.76 \pm 0.07(stat.)^{+0.12}_{-0.11}(syst.)\pm 0.05(theor.) )\times10^6 cm^{-2}s^{-1} $

$\phi(\nu_\mu,\nu_\tau)=( 3.41^{+0.45 +0.48}_{-0.45 -0.45} )\times10^6 cm^{-2}s^{-1}$

and calculated in this work elastic neutrino-electron
cross-sections

$\sigma^{B-8}_{\nu_e} = 5.96\times10^{-44} cm^2$,

$\sigma^{B-8}_{\nu_{\mu,\tau}} = 7.83\times10^{-45} cm^2$.

We obtain the expected number of events for 1 ton of $^{76}Ge$
target ( isotopically enriched to  86 \% in $^{76}Ge$ ) per year

$R^{B-8}_{\nu_e}$ = 0.84 events/( t y ),

$R^{B-8}_{\nu_{\mu,\tau}}$ = 0.22 events/ (t y).

Calculation of count rate for pp and $^7Be$ neutrinos

$R^{pp}$ = 581 events/(t y )

$R^{Be-7}$ = 229 events/(t y )

for fluxes from the Solar Standard Model~\cite{bahc98}

$\phi^{pp}(\nu_e)=( 5.94 \pm 0.01 )\times10^{10} cm^{-2}s^{-1} $

$\phi^{Be-7}(\nu_e)=( 0.48 \pm 0.09 )\times10^{10} cm^{-2}s^{-1}$

have been performed and then compared  with those ones
from~\cite{genius}

$R^{pp}$ = 658 events/(t y )

$R^{Be-7}$ = 219 events/(t y ).

Since pp neutrinos have a continuous energy spectrum with endpoint
at 0.42 MeV and $^7Be$ neutrino from $ ^7Be + e^- \rightarrow ^7Li
+ \nu_e + \gamma$ reaction yields a monoenergetic line at 0.862
MeV, elastic scattering of these neutrinos can not give events
into ROI.

The experiments under consideration are supposed to consist of an
array of about 300 HPGe detectors, with mass $\approx $3 kg each,
and the sensitive volume divided into 12 separate cells to
discriminate background since neutrino interaction is taking place
in a single cell.

 Recoil electron energy deposition for these
experiments for a single Ge detector was performed with GEANT 3.21
package~\cite{geant}, as well as background caused by scattering
of $^8B$ solar neutrino on electrons of Ge detector in ROI is
$B_{ROI}=4.6\times10^{-4}$ counts/(keV$\cdot t\cdot$ y).

 In the experiments HPGe detectors are placed in some cool medium, such as liquid nitrogen or liquid
argon, or copper. The medium can contribute to background
 due to recoil electrons from solar neutrino electron scattering in the media.
Calculations for liquid nitrogen to be used in GENIUS experiment
were done.  The background is 4.5 \% of that of Ge detectors given
above.

Antineutrino flux from nuclear power plant reactors being
$\phi_{\tilde\nu_e} \approx 4\times10^6 \tilde\nu_e\times cm^{-2} \times sec^{-1}$
yields background of $2.3\times10^{-5}$
counts/(keV$\cdot t\cdot$y), which is 11\% of $B_{ROI}$.

(The antineutrino flux value taken above is that of the KamLAND
experiment's place).

Calculations to define contribution due to geo antineutrinos was
estimated too, it is  about 0.9 \% of $B_{ROI}$~\cite{geo}.

Similar calculations for liquid xenon 10 tons detector  were performed.

The count rate of ES events could be given

 \begin{equation}
{ R_{ES} \approx ( 1.1\times10^{-3}/ (keV\cdot t\cdot y) ) ( ROI \frac{ M}{M_{mol}} N_e )}
\label{eq:escr}
\end{equation}

M - is mass of detector in tons, $M_{mol}$ - is target molecular
weight, ROI is energy region width in keV, $N_e$ - is number of
electrons in target molecule.

For 400 kg of $^{136}Xe$, neutrinoless double beta decay  with $T^{0\nu}_{1/2}=4.\times10^{27}$ y
will give $\approx$ 0.3 counts per year.
From formula (~\ref{eq:escr} )  for 10 tons xenon detector  the  ES count rate is same.

As EXO project cannot be able to reconstruct the tracks of the
emitted electrons, the background due to ES will not be rejected.
So the sensitivity is $ T^{0\nu}_{1/2} \approx 4.\times10^{27}$ years.

Using~(\ref{halflife}) and taking  nuclear structure factor $ F_N
= G^{0\nu}\times M_{0\nu}^2 = 1.18\times10^{-13} y^{-1}$
from~\cite{nme} we can obtain sensitivity in terms of effective
majorana neutrino mass:

\begin{equation}
 \mid < m_{\nu_e}>\mid = \frac{m_e}{\sqrt{F_N\times T^{0\nu\beta\beta}_{1/2} }} = \ 23.5 \ meV.
\end{equation}

In Fig.1, one can see three energy spectra  in the ROI, namely,
the two electron sum spectrum of $^{136}Xe$  $( 2\beta2\nu )^{136}Ba$ decay
with $T^{2\nu}_{1/2}=2.0\times10^{22}$ y; the
calculated recoil electron spectrum $ \nu + e^- \rightarrow \nu +
e^-$  for neutrinos from $^8B  \rightarrow 2\alpha + e^+ + \nu_e $
solar reaction, and the third one is the expected peak,
corresponding to $^{136}Xe(0\nu\beta\beta)^{136}Ba$ decay with
$T^{0\nu}_{1/2}=4.0\times10^{27}$ y and FWHM(2480 keV) = 60.0 keV.

Similar situation is shown for case $^{76}Ge$ with $T^{2\nu}_{1/2}=1.74\times10^{21}$y~\cite{klap}
and FWHM(2039 keV) = 5.0 keV in Fig2. But in this case, the contribution of ES events
is only 2\%.

To summarize, for the planned experiments with detectors enriched in
$^{76}Ge$ with mass upto $\approx$ 10 tons, the background from elastic scattering of
solar neutrinos could be considered as negligible.

For experiments with mass of $\approx$ 10 \ tons and  resolution of tens keV,
such as EXO or XMASS, this background channel though by its
universal character restricts sensitivity of double beta decay
experiments at a level of $T^{0\nu\beta\beta}_{1/2} \approx
4.\times10^{27}$ y or the same, in terms of effective majorana
mass, $ \mid < m_{\nu_e}>\mid \approx 23.5 \ meV$.

\clearpage \onecolumn
\begin{figure}[htbp]
\centering
\includegraphics*[bb=.1in .1in 9in 7.3in,scale=0.95]{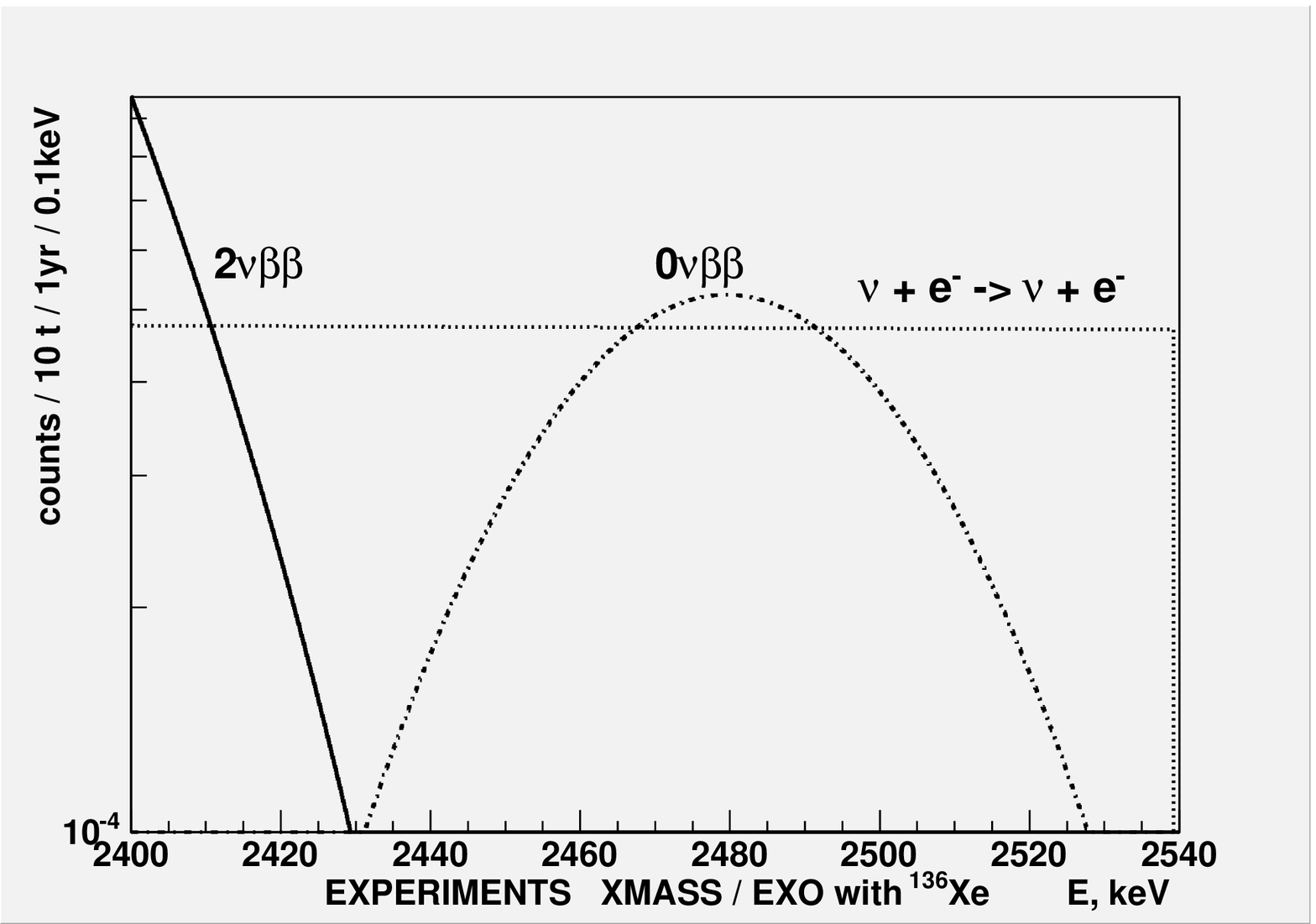}
\caption[1]{
{\bf Fig.1. } Spectra in the vicinity of ROI of $^{136}Xe(0\nu\beta\beta)^{136}Ba$ decay.
Solid line is two electron energy sum spectrum of $^{136}Xe(2\beta2\nu )^{136}Ba$ decay  with $T^{2\nu}_{1/2}=2.0\times10^{22}$y. Dotted line is calculated recoil electron spectrum \mbox {$ \nu + e^- \rightarrow \nu + e^-$}  for neutrinos from $^8B  \rightarrow 2\alpha + e^+ + \nu_e $ solar reaction. Dash-dot line is expected peak, corresponding to $^{136}Xe(0\nu\beta\beta)^{136}Ba$ decay with  
$T^{0\nu}_{1/2}=4.0\times10^{27}$ y and FWHM(2480 keV) = 60.0 keV. }
\end{figure}

\begin{figure}[htbp]
\centering
\includegraphics*[bb=.1in .1in 9in 5.3in,scale=0.95]{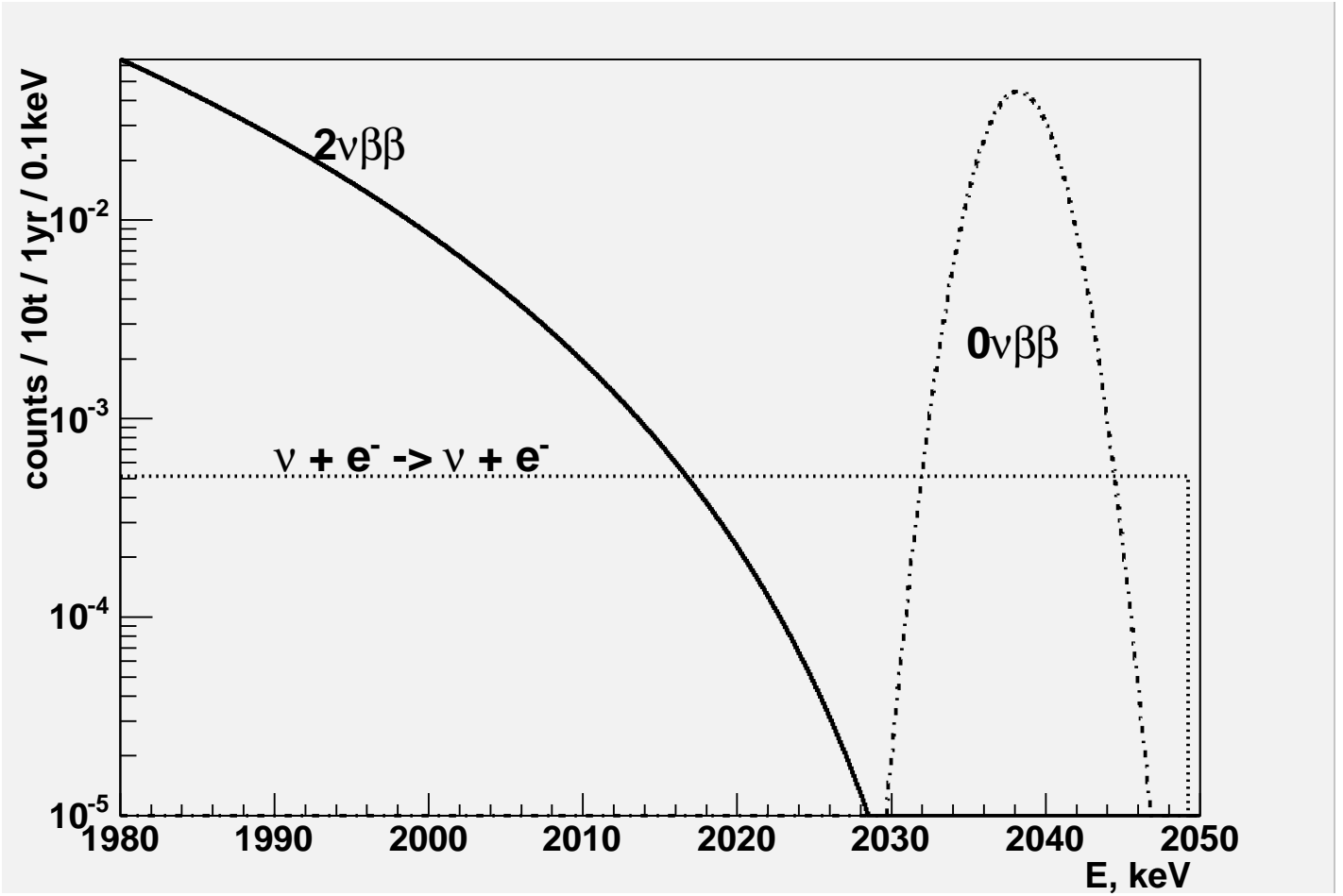}
\caption[2]{
{\bf Fig.2. } Spectra in the vicinity of ROI of $^{76}Ge(0\nu\beta\beta)^{76}Se$ decay.
Solid line is two electron energy sum spectrum of $^{76}Ge(2\beta2\nu )^{76}Se$ decay  with
$T^{2\nu}_{1/2}=1.74\times10^{21}$ y. Dotted line is
calculated recoil electron spectrum \mbox {$ \nu + e^- \rightarrow
\nu + e^-$}  for neutrinos from $^8B  \rightarrow 2\alpha + e^+ +
\nu_e $ solar reaction. Dash-dot line is expected peak,
corresponding to $^{76}Ge(0\nu\beta\beta)^{76}Se$ decay with
$T^{0\nu}_{1/2} = 4.2 \times 10^{27}$ y and FWHM(2039 keV) = 5.0 keV. }
\end{figure}

\end{document}